\documentclass[12pt]{article}
\usepackage{epsfig,fullpage}
\newcommand{\lapprox}{\raisebox{-0.5ex}{$\
\stackrel{\textstyle<}{\textstyle\sim}\ $}}
\newcommand{\gapprox}{\raisebox{-0.5ex}{$\
\stackrel{\textstyle>}{\textstyle\sim}\ $}}
\begin{document}
\begin{center}

\begin{flushright}
     SWAT/01/308 \\
\end{flushright}
\par \vskip 10mm

\vskip 1.2in

\begin{center}
{\LARGE\bf
The (2+1)-dimensional Gross-Neveu model 
\vskip 0.12in
with a $U(1)$ chiral symmetry at
\vskip 0.12in
non-zero temperature}

\vskip 0.7in
S.J. Hands $\!^a$, J.B. Kogut $\! ^b$ and
 C.G. Strouthos $\!^{a}$ \\
\vskip 0.2in

$^a\,${\it Department of Physics, University of Wales Swansea,\\
Singleton Park, Swansea, SA2 8PP, U.K.} \\
$^b\,${\it Department of Physics, University of Illinois at Urbana-Champaign,\\
Urbana, Illinois 61801-3080, U.S.A.}\\
\end{center}

\vskip 1.0in 
{\large\bf Abstract}
\end{center}
\noindent
We present results from numerical simulations of the $(2+1)$-dimensional 
Gross-Neveu model with a  
$U(1)$ chiral symmetry and $N_f=4$ fermion species at non-zero temperature.
We provide evidence that there are two different chirally symmetric phases,
one critical and one with finite correlation length, 
separated by a Berezinskii-Kosterlitz-Thouless transition. 
We have also identified a regime above the critical temperature
in which the fermions acquire a screening mass even in the absence of chiral
symmetry breaking, analogous to the pseudogap behaviour observed in cuprate
superconductors.

\newpage

\section{Introduction}

Phase transitions in QCD and model field theories have been studied intensively  over the last 
decade both analytically and numerically. Understanding the phase diagram of QCD is becoming 
increasingly important in view of recent experimental efforts to create and detect the
quark-gluon plasma in the relativistic heavy ion collisions at BNL and CERN.

Since the problem of chiral symmetry breaking and its restoration is intrinsically 
non-perturbative the number of available techniques is limited and most of our knowledge
about the phenomenon comes from lattice simulations. Because of the complexity of QCD
with dynamical fermions, studies so far have been done on lattices with modest size and 
in various cases the results are distorted by finite size and discretization effects. 
The three-dimensional Gross-Neveu model (GNM$_3$) has been proved to be an interesting 
and tractable model to study chiral phase transitions both numerically by means of lattice simulations
and analytically in the form of large-$N_f$ expansions [1$-$10].

In this letter we present results of numerical simulations of the $U(N_f)_V$-invariant GNM$_3$ with a 
$U(1)$ chiral symmetry at non-zero temperature. 
This model is described by the 
following continuum Lagrangian density (we work in Euclidean space throughout):
\begin{equation}
{\cal L}= \bar{\Psi}_i(\partial\hskip -.5em / + m_0) \Psi_i
- \frac{g^2}{2 N_f} [(\bar{\Psi}_i \Psi_i)^2 - (\bar{\Psi}_i \gamma_5 \Psi_i)^2].
\end{equation}
We treat $\Psi_i$, $\bar{\Psi}_i$  as four-component Dirac spinors and the index $i$ runs over $N_f$
fermion species.
In the chiral limit $m_0 \! \rightarrow \! 0$, the $U(1)$ 
invariance is given by
\begin{equation}
\Psi \mapsto e^{i \theta \gamma_5} \Psi\;\;\;;\;\;\;
\bar\Psi\mapsto\bar\Psi e^{i\theta\gamma_5}. 
\end{equation}
This Lagrangian is considerably easier to treat, both numerically and analytically, 
if it is bosonized by introducing auxiliary fields $\sigma$ and $\pi$:
\begin{equation}
{\cal L}= \bar{\Psi}_i(\partial\hskip -.5em / + m_0 + \sigma + i \gamma_5 \pi)\Psi_i
+ \frac{N_f}{2 g^{2}} (\sigma^{2}+ \pi^2).
\end{equation}
At tree level, the fields $\sigma$ and $\pi$ have no dynamics; they are truly 
auxiliary fields. However, they acquire dynamical content by dint of quantum effects
arising from integrating out the fermions. 
The model is renormalizable in the $1/N_f$ expansion unlike in the loop 
expansion \cite{rosen91}.

Apart from the obvious numerical advantages of working with a relatively simple model in a reduced
dimensionality
there are several other motivations for studying such a model: (i) at $T\!=\!0$ for sufficiently 
strong coupling it exhibits spontaneous chiral symmetry breaking 
and the pion field $\pi$ 
is the associated Goldstone boson; (ii) the spectrum 
of excitations contains both baryons and mesons, 
i.e. the elementary fermions and
the composite fermion$-$anti-fermion states; (iii) the model has an interacting continuum limit
for a critical value of the coupling $1/g_c^2$, which has a numerical value 
$\approx1.0/a$ in the large-$N_f$ limit if a lattice
regularisation is employed \cite{kogut93}; (iv) numerical simulations of the 
model with chemical potential $\mu\not=0$ show qualitatively correct behaviour, 
unlike QCD
simulations \cite{general,hands}.

The phase diagram of GNM$_3$ with various global symmetries
at non-zero temperature and density 
has been studied extensively in [3$-$10]. More specifically \cite{strouthos,cox}, 
it was shown that the thermally induced phase
transition of the $Z_2$-symmetric model belongs to the two-dimensional Ising universality class
in accordance with the dimensional reduction scenario \cite{wilczek} which predicts
that the long-range behaviour at the chiral phase transition is that of the 
$(d-1)$ spin model with the same symmetry, because the IR region of the 
system is dominated by 
the zero Matsubara mode of the bosonic field. 
Outside the non-trivial region the critical properties
of the system are described by mean field theory according to the 
large-$N_f$ prediction. 
It was also shown that the two-dimensional Ising scaling region is 
suppressed by a factor $N_f^{-1/2}$ \cite{strouthos}.

At leading order in $1/N_f$ the $U(1)$ model has a second order chiral phase 
transition at $T_c = \frac{m_f}{2 \ln2}$ \cite{rosen91}, 
where $m_f$ is the fermion dynamical mass at 
zero temperature. 
The leading order effective potential has the same form as the discrete symmetry case with the replacement
$\sigma^2 \! \rightarrow \! \sigma^2 \! + \! \pi^2$.
This conclusion is expected to be valid only when $N_f$ is strictly infinite, i.e. when the fluctuations of the 
bosonic fields are neglected, since otherwise it runs foul of
the Coleman-Mermin-Wagner (CMW) theorem \cite{cmw}, which states that in two-dimensional systems  
the continuous
chiral symmetry must be manifest for all $T \! > \! 0$. 
Next-to-leading order calculations demonstrated that the symmetry is restored for arbitrarily small
$T \! < \! 0$ [7$-$10].
As argued in \cite{appelquist,babaev} 
the model is expected in accordance with the dimensional reduction scenario 
to undergo a Berezinskii-Kosterlitz-Thouless (BKT) transition \cite{kosterlitz} 
at $T_{BKT}$ which is associated with the
unbinding of vortices like in the two-dimensional $XY$ model.
In terms of the reduced temperature $t \equiv T-T_{BKT}$ the scaling behaviour of the correlation 
length, susceptibility and the specific heat is given \cite{kosterlitz} by
\begin{equation}
\xi(t) \sim e^{at^{-\nu}}\!, \;\;\; \chi(t) \sim \xi^{2-\eta}\!, \;\;\; C_v \sim \xi^{\alpha/\nu}+\mathrm{constant},
\end{equation}
where for $t \rightarrow 0^+$, $\nu=1/2$, $\eta=1/4$ and $\alpha=-d\nu=-1$.
It is easier to visualize this scenario if we use the ``modulus-phase'' parametrization 
$\sigma + i \pi \equiv \rho e^{i \theta}$.
In two spatial dimensions logarithmically divergent infrared fluctuations do not allow the phase 
$\theta$ to take a fixed direction and therefore prevent spontaneous symmetry
breaking via $\langle\theta\rangle\not=0$.
The critical temperature $T_{BKT}$ is expected to separate two \emph{different} chirally symmetric phases:
a low $T$ phase, which is characterized by power law phase correlations 
$\langle e^{i\theta(x)} e^{-i\theta(0)} \rangle \sim x^{-\eta(T)}$ at distances $x \gg 1/T$ 
but no long range order (i.e.
a spinwave phase where chiral symmetry is ``almost but not quite broken''), and a high $T$
phase which is characterized by exponentially decaying phase correlations with no long range order.  In 
other words for $0 \leq T \leq T_{BKT}$
there is a line of critical points characterised by a 
continuously varying $0\leq\eta(T)\leq{1\over4}$. 
However, since the amplitude $\rho$ is neutral under $U(1)$, 
the dynamics of the low temperature phase do not preclude the
generation of a fermion mass $m_f\propto\rho$ 
whose value may be comparable with the naive
prediction of the large-$N_f$ approach; this was demonstrated 
in $d=(1+1)$ in \cite{witten},
in which results from an exactly soluble fermionic model \cite{sinclair}
are generalised to GNM$_2$. The effect may be understood in
terms of the spectral function $\rho(s)$. For orthodox mass generation,
the spectral function in the broken chiral symmetry phase is a
simple pole $\rho(s)\propto\delta(s-m_f^2)$. By contrast in the BKT scenario 
\cite{witten} the function is 
modified to a branch cut $\rho(s)\propto(\surd s-m_f)^{\eta-1}
\theta(\surd s-m_f)$;
although chiral symmetry is manifest, the propagating fermion constantly
emits and absorbs massless scalars and hence has indefinite chirality.

In \cite{babaev} it is 
shown that in GNM$_3$ with $T>0$, 
next-to-leading order corrections cause fermion mass generation 
to occur for $T<T_*$, where $T_*>T_{BKT}$,
ie. it predicts a ``pseudogap'' phase in which the system is non-critical, 
chiral symmetry is manifest, and yet the energy required to create a
spin-${1\over2}$ excitation does not vanish. Such behaviour is interesting
because it is characteristic of the underdoped phase of certain cuprate
superconductors \cite{randeria}.

We have attempted to check the validity of 
(a) the BKT scenario, and (b) mass generation without 
chiral symmetry breaking, by performing lattice simulations. 
Our results with $N_f=4$ 
show that the symmetry is restored for a 
large range of non-zero temperatures. 
The susceptibility of the order parameter diverges 
at a temperature which we associate with $T_{BKT}$.
We also verify the prediction of \cite{babaev} 
that the fermions have non-zero mass in a region where the system is not
critical, i.e. for $T>T_{BKT}$. Our simulations have not shown evidence for 
a second phase transition to a massless phase 
at $T=T_*$, as predicted by large-$N_f$ methods \cite{babaev}, although
we cannot rule it out.

\section{Numerical Simulations}
The model in its bosonised form can be formulated on the lattice using the following action:
\begin{equation}
S_{lat}= \sum_{x,y} \bar{\chi}(x) M_{x,y} \chi(y)
+\frac{1}{8} \sum_{x} \bar{\chi}(x) \chi(x)  [ \sum_{ \langle \tilde{x},x \rangle}
\sigma(\tilde{x}) + i \epsilon(x) \sum_{ \langle \tilde{x},x \rangle}
\pi(\tilde{x}) ]
+ \frac{1}{2g^{2}} \sum_{\tilde{x}} [\sigma^2(\tilde{x}) + \pi^2(\tilde{x})],
\label{gnaction_lattice}
\end{equation}
where $\chi$ and $\bar{\chi}$ are Grassmann-valued 
staggered fermion fields
defined on the lattice sites, the auxiliary field $\sigma$ is defined on the dual lattice
sites and the symbol $\langle \tilde{x},x \rangle$ denotes the set of 8 dual lattice
sites $\tilde{x}$ surrounding the direct lattice site $x$
\cite{kogut93}. The fermion kinetic operator $ M $ is given by
\begin{equation}
M_{x,y} = \frac{1}{2} \left[ \delta_{y,x+\hat{0}} - \delta_{y,x-\hat{0}} \right]
+ \frac{1}{2} \sum_{\nu=1,2} \eta_{\nu}(x) \left[ \delta_{y,x+\hat{\nu}} -
\delta_{y,x-\hat{\nu}} \right],
\end{equation}
where $\eta_{\nu}(x)$ are the Kawamoto-Smit phases $(-1)^{x_0+...+x_{\nu-1}}$
and the symbol $\epsilon(x)$ denotes the alternating phase $(-1)^{x_0+x_1+x_2}$.
The simulations were performed by using the standard hybrid Monte Carlo algorithm 
in which complex bosonic pseudofermion fields $\Phi$ are updated using the action 
$\Phi^{\dagger}(M^{\dagger}M)^{-1} \Phi$.
According to the discussion in \cite{hands}, simulation of $N$ staggered
fermions describes $N_f=4N$ continuum species; 
the full symmetry of the lattice model in the
continuum limit, however, is $U(N_f/2)_V \otimes U(N_f/2)_V \otimes U(1)$ 
rather than $U(N_f)_V \otimes U(1)$. 
At non-zero lattice spacing the symmetry group is smaller: $U(N_f/4)_V \otimes
U(N_f/4)_V \otimes U(1)$. In this study we have used $N=1$.

\begin{figure}[p]

                \centerline{ \epsfysize=3.6in
                             \epsfbox{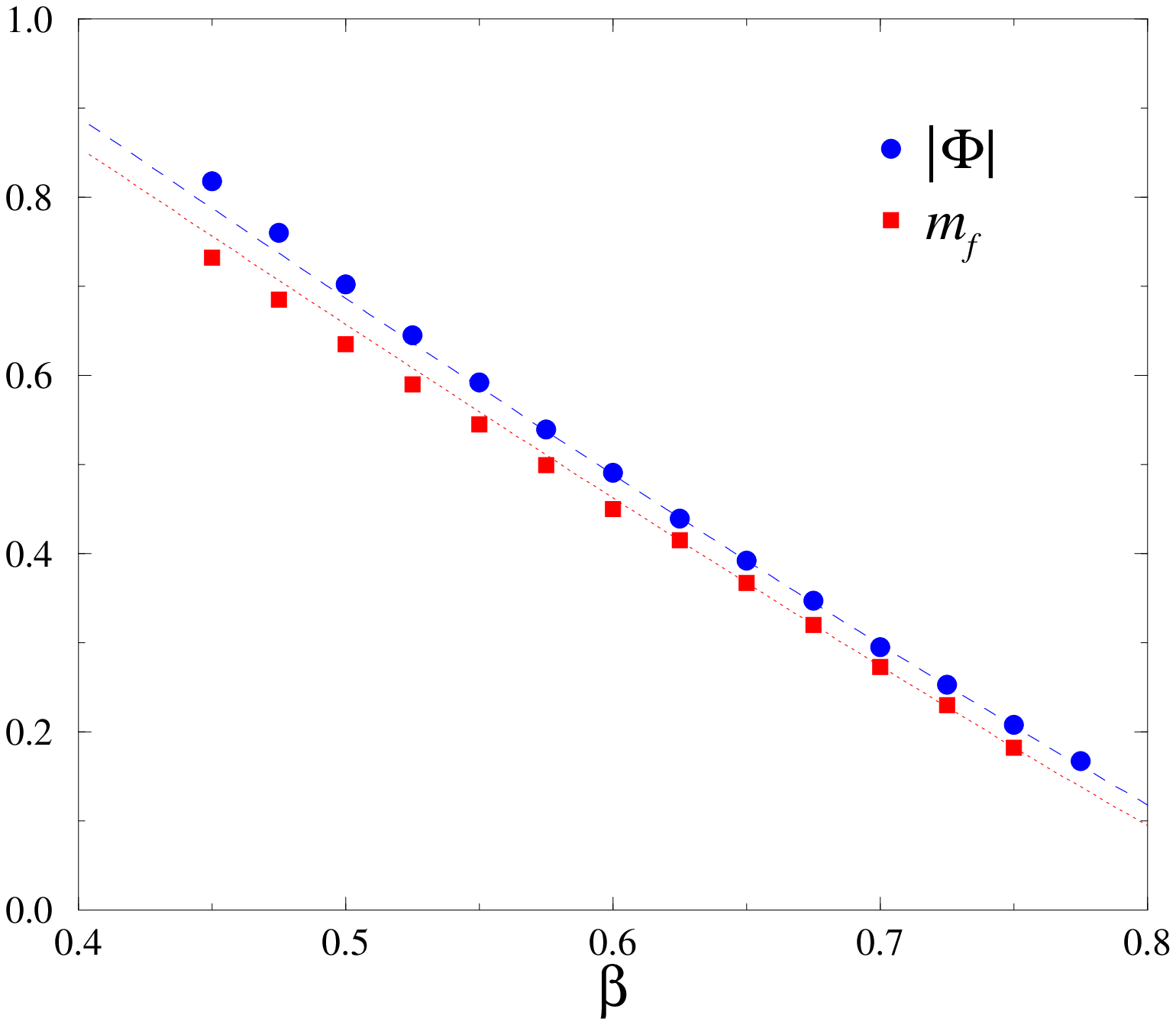}}

\smallskip
\caption[]{Order parameter $\vert\Phi\vert$ and fermion mass $m_f$
at $T=0$ vs. coupling $\beta$ on a $24^2\times36$ lattice.}
\label{fig:t=0}
\end{figure}

In order to study the behaviour of the chiral symmetry first at $T=0$ and then for $T>0$ we set the fermion bare mass
to zero. 
Without the benefit of this symmetry breaking interaction, the direction 
of symmetry breaking changes over the course of the run so that
$\Sigma \equiv \frac{1}{V} \sum_x \sigma(x)$ and 
$ \Pi \equiv \frac{1}{V} \sum_x \pi(x) $
average to zero over the ensemble. It is in this way that the absence of spontaneous breaking 
of this continuous symmetry on a finite lattice is enforced. 
The next best thing to measure is
an effective ``order parameter'' $|\Phi| \equiv \sqrt{\Sigma^2 + \Pi^2}$, which is a projection onto 
the direction of $\Phi^{\alpha}  \equiv (\Sigma, \Pi)$ separately for each configuration \cite{hasenfratz}.  
In the case of chiral symmetry breaking $|\Phi|$ differs from the true order parameter 
$\langle\Sigma\rangle$ extrapolated to the chiral limit, because in the absence of
a symmetry breaking term it is impossible to disentagle the fluctuations of the order parameter 
field from those of the Goldstone modes. Including the effects of the latter will give 
$|\Phi|>\Sigma_0$, where $\Sigma_0$ denotes the value of $\Sigma$ in the chiral limit.
The simulations at $T=0$ were performed on $24^2\times36$ lattices for
$\beta\equiv1/g^2=0.45$ -- 0.775.
In Fig.\ref{fig:t=0} we plot $|\Phi|$ and the fermion mass $m_f$ versus $\beta$. 
We fitted these two observables to the scaling functions 
$|\Phi|=a_1(\beta^{bulk}_c-\beta)^{\beta_{mag}}$ and 
$m_f=a_2(\beta^{bulk}_c-\beta)^{\nu}$ for $\beta=0.625-0.750$. 
We extract $\nu=1.08(5)$, $\beta_{mag}=1.1(1)$ 
which are consistent with the values of the 
large-$N_f$ prediction $\beta=\nu=1$ \cite{kogut93}. The values of the 
associated bulk critical couplings are:
$\beta^{bulk}_c=0.87(3)$ from the order parameter fit and 
$\beta^{bulk}_c=0.86(1)$ from the fermion mass fit.
At very strong coupling (small $\beta$) the data deviate slightly 
from the fitted scaling functions due to discretization 
effects. 

\begin{figure}[p]

                \centerline{ \epsfysize=3.6in
                             \epsfbox{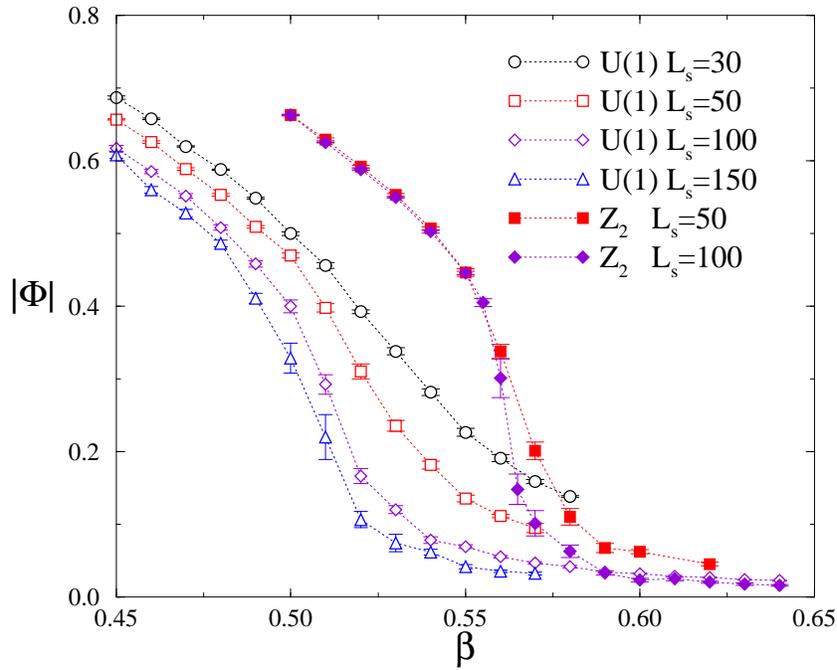}}

\smallskip
\caption[]{Order parameter $\vert\Phi\vert$ vs. $\beta$ for $L_t=4$
for both $U(1)$ and $Z_2$ symmetric models.}
\label{fig:order_para}
\end{figure}

In order to study the effects of non-zero temperature we performed simulations 
on asymmetric lattices with 
constant temporal extent $L_t=4$ and spatial extents
$L_s=30,50,100,150$. We vary $T$ by changing $\beta$. At $\beta^{bulk}_c$ 
the lattice spacing 
becomes zero and $T \rightarrow \infty$. 
Simulations of the $Z_2$ model with 
$L_t=4$ show a thermally induced phase transition at 
$\beta^{Z_2}_c=0.565(3)$, which should be compared with the 
value $\beta_c=0.76$ expected in the $N_f\to\infty$ limit on a lattice with 
$L_t=4$.
For $L_s=100$ and $150$ we accumulated approximately $30,000$ to $45,000$ 
trajectories for the range of
couplings $\beta \leq 0.54$ and $10,000$ to $20,000$ for larger values of $\beta$. 
The 
trajectory length was approximately $1.5$ and was chosen at random from a Poisson 
distribution in order to decrease autocorrelation times. For $L_s=30$ and $150$ 
we chose a trajectory length of approximately $1.0$ and accumulated
$40,000$ to $50,000$ trajectories for $\beta \leq 0.54$ and 
$20,000$ to $30,000$ for larger values of $\beta$.
In Fig.\ref{fig:order_para} we plot $|\Phi|$ versus $\beta$
for the various lattice sizes together with the results from simulations
of the $Z_2$ model on lattice sizes $4 \times 50^2$ and $4 \times 100^2$. It is clear that 
the order parameter of the $Z_2$ model is independent of the lattice size until just before the
transition at $\beta=\beta_c^{Z_2}$, 
whereas in the $U(1)$ model $|\Phi|$ has a strong size dependence for a large range of values 
of $\beta$, i.e. it
decreases rapidly as the spatial volume increases in accordance with the expectation that chiral 
symmetry should be restored for $T>0$.
We believe that the signal for chiral symmetry breaking at very low $T$ is a result of finite size corrections. 
The finite spatial extent $L_s$ provides a cut-off for the divergent correlation length and according to 
the BKT scenario 
the slow decay of the correlation function $\langle e^{i\theta(x)} e^{-i\theta(0)} \rangle$ 
with exponent $\eta(T)<0.25$ for $T<T_{BKT}$ ensures a non-zero magnetization even in a system with very large size.
In other words on small lattices the spinwave phase is expected to look like a broken phase.
A similar effect has been observed in the two-dimensional Gross-Neveu model with a discrete symmetry
at non-zero $T$ \cite{wyld}: For large-$N_f$ the density of kinks (whose condensation is responsible 
for the restoration of chiral symmetry at $T>0$) is exponentially suppressed and therefore on lattices 
with spatial extent smaller than the kink size the system looks as if it is in the broken phase.

\begin{figure}[p]

                \centerline{ \epsfysize=3.6in
                             \epsfbox{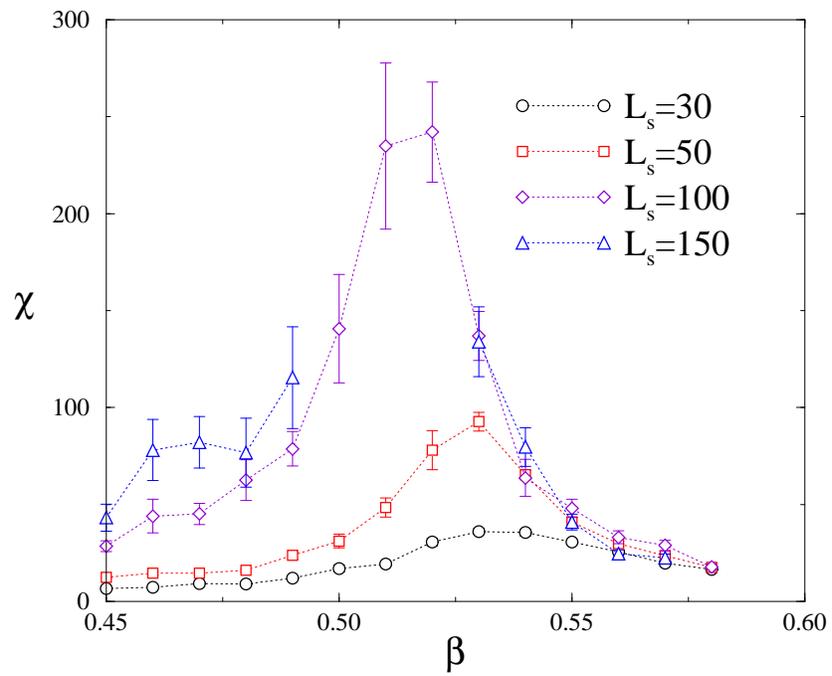}}

\smallskip

\caption[]{Susceptibility $\chi$ vs. $\beta$.}
\label{fig:suscept}
\end{figure}

\begin{figure}[p]

                \centerline{ \epsfysize=3.6in
                             \epsfbox{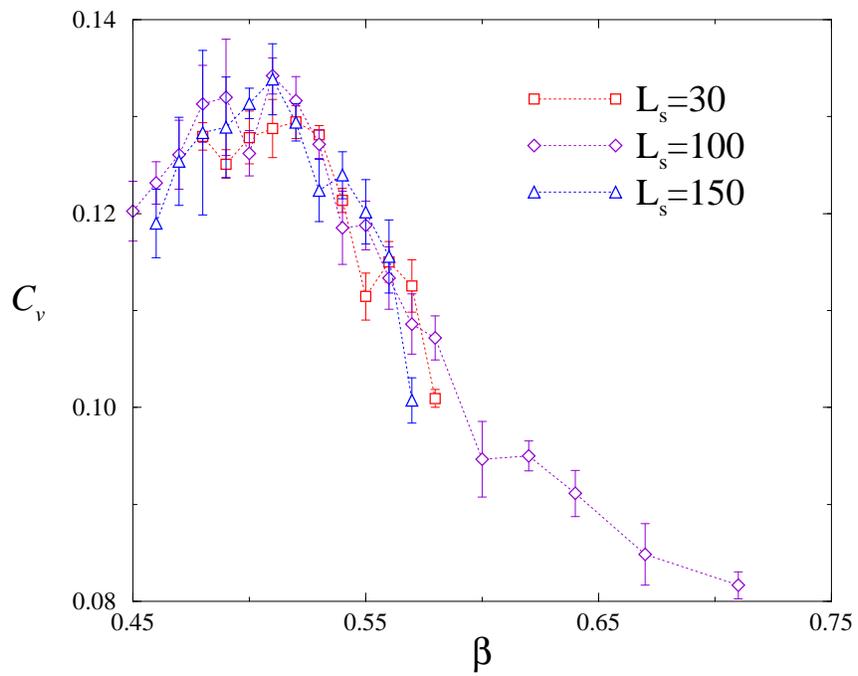}}

\smallskip

\caption[]{Specific heat $C_v$ vs. $\beta$.}
\label{fig:spec_heat}
\end{figure}

In Fig.\ref{fig:suscept} we plot the susceptibility of the order parameter 
\begin{equation}
\chi = V (\langle |\Phi|^2 \rangle - \langle |\Phi| \rangle^2),
\end{equation}
measured on lattices with spatial size $L_s=30,50,100,150$. It is clear that $\chi$ has a peak 
for $L_s=30,50,100$  at $\beta_{BKT}(L_s) \simeq 0.54, 0.53, 0.515$ 
respectively.  
The values of $\chi$ near the $L_s=150$ peak are very noisy and we don't
plot them in this graph.
This is clear evidence of a phase transition, with a 
critical coupling $\beta_{BKT}$ significantly less
than $\beta_c^{Z_2}$.
The transition occurs at a
much lower temperature than the critical 
temperature of the GNM$_3$ with a $Z_2$ symmetry, 
because the infrared fluctuations are stronger in the continuous symmetry case.
Another interesting observation is that in 
the low temperature phase the susceptibility has  stronger
size dependence than in the high temperature phase 
and the error bars at low $T$ are much larger that at high $T$ despite the fact that the statistics
at low $T$ are much larger than at high $T$.
This is evidence that the system is critical 
in the low $T$ phase in accordance 
with the BKT scenario. 
Another useful quantity to measure is the so-called specific heat $C_v$, 
which we calculated from the
fluctuations of the bosonic action 
$S_b = \frac{1}{2} \sum_x [\sigma^2(x) + \pi^2(x)]$.
$C_v$ is given by
\begin{equation}
C_v = \frac{\beta^2}{V} (\langle S_b^2 \rangle - \langle S_b \rangle^2),
\end{equation}
and is plotted versus coupling for $L_s=50,100,150$ in
Fig.\ref{fig:spec_heat}.
It is clear that (a) $C_v$ has a broad peak at $\beta \simeq 0.50$ and (b)
it does not show any divergent behaviour or significant finite size effects
which is consistent with the BKT scenario, according to which $\alpha=-2$.
Analogous behaviour has been observed in numerical studies of the 
$XY$ model \cite{xy}.

\begin{figure}[p]

                \centerline{ \epsfysize=3.6in
                             \epsfbox{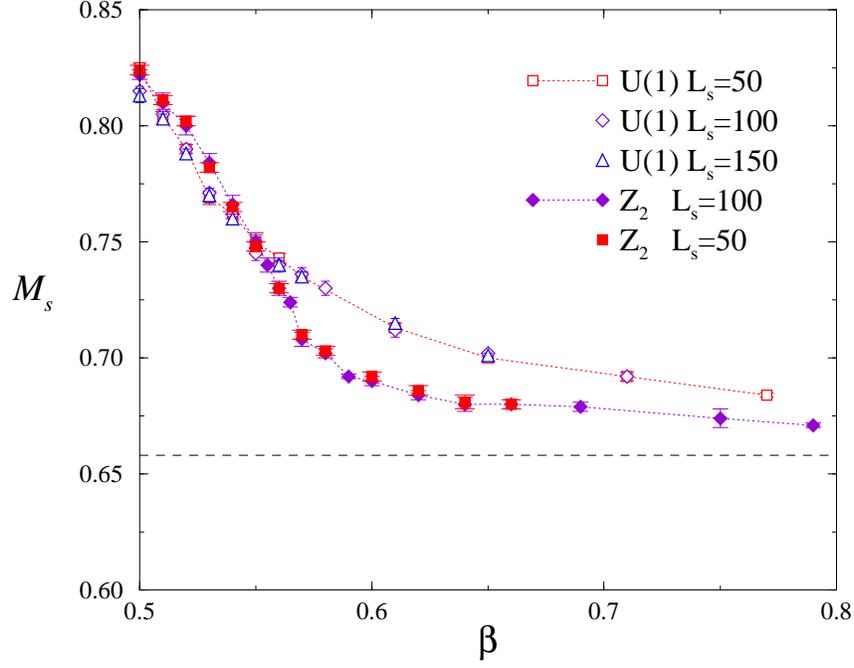}}

\smallskip

\caption[]{Fermion screening mass $M_s$ vs. $\beta$. The horizontal line 
shows the lowest Matsubara mode $M_0^{lat}$.}
\label{fig:mass}
\end{figure}

\begin{figure}[p]

                \centerline{ \epsfysize=3.6in
                             \epsfbox{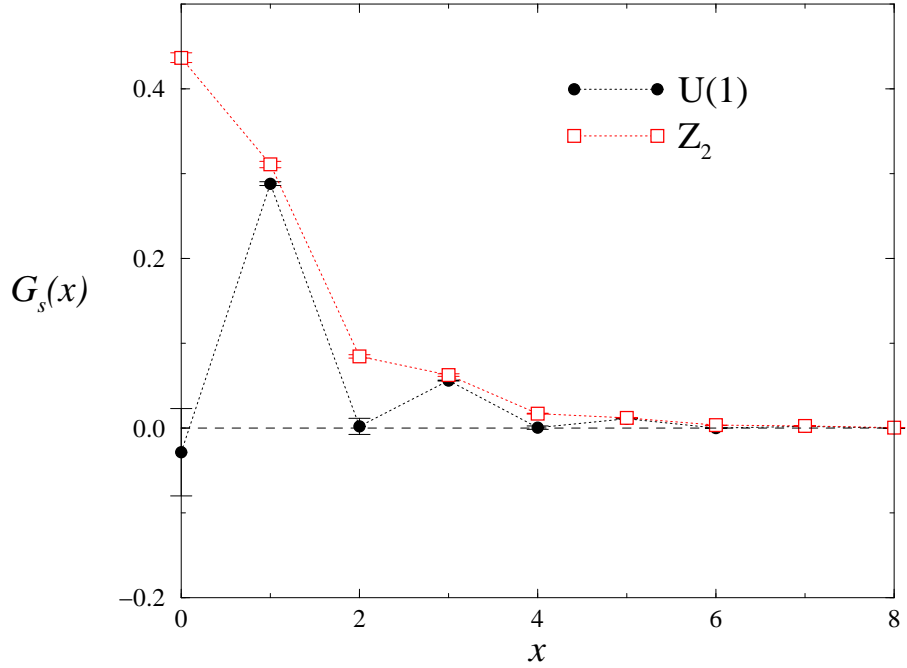}}

\smallskip

\caption[]{Fermion spatial correlators $G_s(x)$ extracted 
from simulations of $U(1)$- and $Z_2$-symmetric GNM$_3$ 
on $4 \times 100^2$ lattices at 
$\beta =0.51$.}
\label{fig:ferm_prop}
\end{figure}

Now we discuss the issue of fermion mass generation.
Since the temporal direction of our lattices is short, in order to study the 
asymptotic 
behaviour of the fermion propagator we have chosen to measure the spatial
fermion correlator $G_s(x)$ along one of the spatial lattice axes, which 
yields information on spatial correlations in the thermal medium rather than the
spectrum of excitations. The quantity analogous to mass is the
so-called screening mass $M_s$;
for free, massless continuum fermions this quantity is given by 
the lowest Matsubara mode for fermions, ie. $M_s=M_0^{cont} \!=\! \pi T$.  
On a finite temporal lattice this becomes 
$M_0^{lat} \! =\! \sinh^{-1} [\sin(\pi/L_t)] $. 
For $L_t \!=\! 4$
the screening mass in the absence of dynamical mass generation 
is $M_0^{lat} \! \simeq \! 0.658 $. If a dynamical mass $M$ is generated,
then naively we expect
\begin{equation}
M_s^2=M_0^2+M^2.
\label{eq:masses}
\end{equation}
We calculated the screening masses from the exponential decay of the 
spatial correlation functions
using an ansatz for the fit which is motivated by the form of the free 
propagators \cite{fermion}:
\begin{equation}
G_s(x) = A(1-(-1)^x) \sinh(M_s (x-L_s/2)) + B(1+(-1)^x) \cosh(M_s (x-L_s/2)).
\label{eq:prop}
\end{equation}
The screening masses extracted from the simulations 
of GNM$_3$ with $U(1)$ symmetry on lattices with varying $L_s$, 
together with the results of the $Z_2$ model from 
simulations on a $4\times100^2$ lattice, are 
shown in Fig.\ref{fig:mass}. 
It is clear that the fermions have non-zero screening mass $M$ 
in the sense that 
$M_s>M_0$, even when the order parameter
is very small. The screening mass is independent of $L_s$
for almost all values of $\beta$ -- in particular we observe that
$M_s^{U(1)}$ is comparable with $M_s^{Z_2}$ for $\beta<\beta_c^{Z_2}$. In
the $Z_2$ case we ascribe mass generation to orthodox chiral symmetry
breaking. For $U(1)$ the situation is more subtle, as shown in 
Fig.\ref{fig:ferm_prop} where we plot the spatial fermion 
correlator of both models
on a $4 \! \times \! 100^2$
lattice at $\beta=0.51$. The $Z_2$ data are non-vanishing on all timeslices,
consistent with the standard scenario of mass generation via
broken chiral symmetry. By contrast, 
the vanishing of $G_s(x)^{U(1)}$ for even $x$,
corresponding
to $B=0$ in (\ref{eq:prop}), 
signals a manifest chiral symmetry, which follows because
the $U(1)_\epsilon$ symmetry of staggered fermions implies
that the only non-vanishing elements of the propagator are $G_{eo}$ and
$G_{oe}$.
By comparing Fig.\ref{fig:suscept} with Fig.\ref{fig:mass} we infer 
that the fermions are massive, ie. $M>0$,  
in the phase where both chiral symmetry is restored and
the system is not critical, 
ie. the susceptibility is finite and size 
independent. This implies the existence of a pseudogap phase for
$\beta_{BKT}\lapprox\beta\lapprox\beta_c^{Z_2}$. 

Strictly speaking, in demonstrating that $M_s>M_0$ we have not excluded the
possibility that the relation (\ref{eq:masses}) could be replaced by 
$M_s^2\simeq A(T)M_0^2+M^2$ with $A-1\gg M^2/M_0^2$, 
implying a temperature-dependent modification of the speed
of light rather than mass generation \cite{edwin}.
The similarity of $M_s^{U(1)}$ and
$M_s^{Z_2}$ at low $T$ where $A\approx1$ suggests this is unlikely to be the
dominant effect for $\beta\lapprox\beta_c^{Z_2}$; 
however for $\beta\gapprox0.56$
the two models begin to diverge, and the level ordering switches.
Since the departure from the free-field result in this regime is 
greater for the $U(1)$ model (which has twice as many 
scalar fields) as for the $Z_2$ model, 
it is conceivable that the effect
is due to in-medium modifications of the massless
fermion propagator, and could be calculable in a large-$N_f$ framework.
To understand the physics of this regime the two effects need to be
disentangled via a more detailed study of the dispersion
relation for the spatial correlator, ie. the value of $M_s$ for several
distinct Matsubara modes. This will require simulations closer to the continuum
limit, 
i.e. on lattices with $L_t\geq8$. For this reason we cannot at this stage
identify a second phase transition at $\beta=\beta_*$ where mass generation
switches off. It is possible that such a transition may be an artifact of 
the large-$N_f$ approach \cite{babaev}.

\section{Summary and Outlook}

Our numerical study of the $U(1)$-symmetric GNM$_3$ provided evidence for: 
(a) existence of two chirally symmetric phases separated by a transition 
which is possibly a BKT transition  in 
agreement with the dimensional reduction scenario;
(b) dynamical mass generation  without symmetry breaking in accordance with the prediction 
of the large-$N_f$ calculation and
(c) the existence of a pseudogap phase where the system is not critical (i.e. $T>T_{BKT}$)
and the fermions acquire a non-zero mass.
Our results are another example of what was shown by Witten \cite{witten} in 
GNM$_2$  with a $U(1)$ symmetry
at $T=0$, i.e. that the large-$N_f$ expansion 
is a reliable guide to the properties of the model as long as one 
interprets the results carefully. In \cite{witten} it was shown that although
large-$N_f$ predicts 
spontaneous symmetry breaking which is in contradiction with the CMW theorem, 
the scalar field correlator
falls off like $x^{-1/N_f}$, 
i.e. there is an ``almost long range order'' together with fermion mass
generation which does not break chiral symmetry because the physical fermion is a superposition of
positive and negative chirality states and has zero net chirality.
The results of \cite{strouthos},
where it was shown that for GNM$_3$ with 
a $Z_2$ chiral symmetry at non-zero $T$
the region (in the space of $m/T$) of the large-$N_f$ 
behaviour squeezes out the two-dimensional 
Ising region by a factor $N_f^{-1/2}$, also support the spirit
of Witten's statement.

We are currently extending our work in various directions. We wish

\begin{itemize}

\item to increase the statistics near and below the transition in order
to perform finite size scaling to extract the exponents $\eta(T)$.

\item
to repeat the simulations with $L_t=8$ in order to study the fermion 
dispersion relation.

\item
to study the $N_f$-dependence of our various results, which can be compared
with analytic predictions \cite{babaev}.

\end{itemize}
We also plan to study the  GNM$_3$ with an $SU(2)_L\otimes SU(2)_R$
chiral symmetry at non-zero temperature. 
According to the dimensional reduction scenario
this model's critical properties at non-zero $T$ should be the same as the two-dimensional 
spin model with an $O(4)$ symmetry, i.e. it is expected to have a transition at $T=0^{+}$,
and to have dynamical mass generation for $T>0$.

\section*{Acknowledgements}
Discussions with Egor Babaev and Luigi Scorzato are greatly appreciated.
SJH and CGS were supported by a Leverhulme Trust grant, and also partially by
EU TMR network ERBFMRX-CT97-0122.
JBK was supported in part by NSF grant PHY96-05199. 
The computer simulations were done on the Cray SV1's at NERSC, the Cray T90
at NPACI, 
and on the SGI Origin 2000 at the University of Wales Swansea.  


\newpage

\newpage

\newpage

\end{document}